\font\bbb=msbm10                                                   

\def\R{\hbox{\bbb R}}
\def\Z{\hbox{\bbb Z}}

\def\AP{{\sl Ann.\ Phys.}}
\def\APS{{\sl Acta Physica Slovaca}}

\def\FP{{\sl Found.\ Phys.}}

\def\JMO{{\sl J. Mod.\ Optics}}

\def\PA{{\sl Physica A}}

\def\PLA{{\sl Phys.\ Lett.\ A}}

\def\PPMSJ{{\sl Proc.\ Phys.\ Math.\ Soc.\ Japan}}

\def\PR{{\sl Phys.\ Rev.}}
\def\PRA{{\sl Phys.\ Rev.\ A}}

\def\PRD{{\sl Phys.\ Rev.\ D}}

\def\PRL{{\sl Phys.\ Rev.\ Lett.}}
\def\PRSLA{{\sl Proc.\ Roy.\ Soc.\ Lond.\ A}}

\def\Sc{{\sl Science}}

\def\ZP{{\sl Z. Physik}}
\def\ZPB{{\sl Z. Phys.\ B}}

\def\dajm{\hbox{D. A. Meyer}}

\def\deutsch{\hbox{D. Deutsch}}

\def\zurek{\hbox{W. H. Zurek}}

\def\hfb{\hfil\break}

\catcode`@=11
\newskip\ttglue

   \font\ninerm=cmr9    \font\eightrm=cmr8   \font\sixrm=cmr6
  \font\ninebf=cmbx9   \font\eightbf=cmbx8  \font\sixbf=cmbx6
  \font\nineit=cmti9   \font\eightit=cmti8  
  \font\ninesl=cmsl9   \font\eightsl=cmsl8  
  \font\ninemi=cmmi9   \font\eightmi=cmmi8  \font\sixmi=cmmi6

\font\bigtenbf=cmr10 scaled\magstep2 

\def\ninepoint{\def\rm{\fam0\ninerm}%
  \textfont0=\ninerm \scriptfont0=\sixrm
  \textfont1=\ninemi \scriptfont1=\sixmi
  \textfont\itfam=\nineit  \def\it{\fam\itfam\nineit}%
  \textfont\slfam=\ninesl  \def\sl{\fam\slfam\ninesl}%
  \textfont\bffam=\ninebf  \scriptfont\bffam=\sixbf
    \def\bf{\fam\bffam\ninebf}%
  \tt \ttglue=.5em plus.25em minus.15em
  \normalbaselineskip=11pt
  \setbox\strutbox=\hbox{\vrule height8pt depth3pt width0pt}%
  \normalbaselines\rm}

\def\eightpoint{\def\rm{\fam0\eightrm}%
  \textfont0=\eightrm \scriptfont0=\sixrm
  \textfont1=\eightmi \scriptfont1=\sixmi
  \textfont\itfam=\eightit  \def\it{\fam\itfam\eightit}%
  \textfont\slfam=\eightsl  \def\sl{\fam\slfam\eightsl}%
  \textfont\bffam=\eightbf  \scriptfont\bffam=\sixbf
    \def\bf{\fam\bffam\eightbf}%
  \tt \ttglue=.5em plus.25em minus.15em
  \normalbaselineskip=9pt
  \setbox\strutbox=\hbox{\vrule height7pt depth2pt width0pt}%
  \normalbaselines\rm}

\def\sfootnote#1{\edef\@sf{\spacefactor\the\spacefactor}#1\@sf
      \insert\footins\bgroup\eightpoint
      \interlinepenalty100 \let\par=\endgraf
        \leftskip=0pt \rightskip=0pt
        \splittopskip=10pt plus 1pt minus 1pt \floatingpenalty=20000
        \parskip=0pt\smallskip\item{#1}\bgroup\strut\aftergroup\@foot\let\next}
\skip\footins=12pt plus 2pt minus 2pt
\dimen\footins=30pc

\def\ie{{\it i.e.}}
\def\eg{{\it e.g.}}

\def\@versim#1#2{\lower.5pt\vbox{\baselineskip0pt \lineskip-.5pt
    \ialign{$\m@th#1\hfil##\hfil$\crcr#2\crcr\sim\crcr}}}
\def\gsim{\mathrel{\mathpalette\@versim\>}}

\magnification=1200
\input epsf.tex

\dimen0=\hsize \divide\dimen0 by 13 \dimendef\chasm=0
\dimen1=\hsize \advance\dimen1 by -\chasm \dimendef\usewidth=1
\dimen2=\usewidth \divide\dimen2 by 2 \dimendef\halfwidth=2
\dimen3=\usewidth \divide\dimen3 by 3 \dimendef\thirdwidth=3
\dimen4=\hsize \advance\dimen4 by -\halfwidth \dimendef\secondstart=4
\dimen5=\halfwidth \advance\dimen5 by -10pt \dimendef\indenthalfwidth=5
\dimen6=\thirdwidth \multiply\dimen6 by 2 \dimendef\twothirdswidth=6
\dimen7=\twothirdswidth \divide\dimen7 by 4 \dimendef\qttw=7
\dimen8=\qttw \divide\dimen8 by 4 \dimendef\qqttw=8
\dimen9=\qqttw \divide\dimen9 by 4 \dimendef\qqqttw=9

\parskip=0pt\parindent=0pt

\line{\hfil March 1998}
\line{\hfil quant-ph/9805039}
\bigskip\bigskip\bigskip
\centerline{\bf\bigtenbf SCALE DECOHERENCE IN}
\bigskip
\centerline{\bf\bigtenbf INHOMOGENEOUS POTENTIALS}
\vfill
\centerline{\bf David A. Meyer}
\bigskip 
\centerline{\sl Project in Geometry and Physics}
\centerline{\sl Department of Mathematics}
\centerline{\sl University of California/San Diego}
\centerline{\sl La Jolla, CA 92093-0112}
\centerline{dmeyer@chonji.ucsd.edu}
\smallskip
\centerline{\sl and}
\smallskip
\centerline{\sl Institute for Physical Sciences}
\centerline{\sl Los Alamos, NM}
\vfill
\centerline{ABSTRACT}
\bigskip
\noindent Finite precision measurement factors the Hilbert space of a
quantum system into a tensor product 
$H_{\rm coarse} \otimes H_{\rm fine}$.  This is mathematically
equivalent to the partition into system and environment which forms 
the arena for decoherence, so we describe the consequences of the
inaccessibility of $H_{\rm fine}$ as {\sl scale decoherence}.
Considering the experimentally important case of a harmonic 
oscillator potential as well as a periodic piecewise constant 
potential, we show that scale decoherence occurs for inhomogeneous 
potentials and may explain part of the decoherence observed in recent
and proposed experiments on mesoscopic superpositions of quantum 
states.

\bigskip
\global\setbox1=\hbox{PACS numbers:\enspace}
\global\setbox2=\hbox{PACS numbers:}
\parindent=\wd1
\item{PACS numbers:}  03.65.Bz,  
                      05.30.-d.  
\item{\hbox to \wd2{KEY\hfill WORDS:}}   
                      decoherence; entropy; inhomogeneous potential.

\vfill
\eject

\headline{\ninepoint\it Scale decoherence        \hfil David A. Meyer}
\parskip=10pt
\parindent=20pt

Mesoscopic superpositions of quantum states were prepared [1,2] and
observed to decohere [2] in remarkable experiments reported recently.
These experimental successes coincided with calls for renewed 
theoretical investigation of decoherence [3,4] and proposals for 
experimental measurement of the density matrix [1,5] of mesoscopic 
quantum systems.  Among the motivations for much of this work is the 
prospect of quantum computing [6]:  Not only does the power of quantum 
algorithms depend on quantum superpositions [7,8], but both the ion 
trap [1] and cavity QED [2] systems used in these experiments have 
been proposed for the physical realization of Boolean quantum 
computers [9].  Even these relatively `clean' realizations of quantum 
computation, however, will have runtimes limited by decoherence [10].  
Current proposals for quantum error correction [11] and fault tolerant 
quantum computation [12] aim to circumvent such limits; detailed 
theoretical and experimental understanding of the decoherence of 
quantum states may indicate how to implement these proposals 
successfully.

Careful analysis of the complexity of quantum algorithms considers 
only {\sl finite precision\/} specification of transition amplitudes 
[8], reminding us that only finite precision measurements for quantum
state observables with continuous spectra are physically realistic.  
In this paper we consider finite precision measurement of the position 
of a quantum particle evolving according to the Schr\"odinger equation 
and demonstrate that in the presence of an inhomogeneous potential the
{\sl measured\/} quantum state generically decoheres---independently 
of any coupling to an environment---a phenomenon which we will call 
{\sl scale decoherence}.  The recognition that quantum measurements 
have only finite precision/limited resolution dates back at least to 
von Neumann [13].  Coarse graining of various sorts, including by 
finite precision measurement [14], has more recently been used in the 
consistent histories approach to quantum mechanics to obtain 
decoherent sets of histories [15].  The scale decoherence we exhibit 
here in the reduced density matrix is the Schr\"odinger picture 
version of this phenomenon and should be apparent in experimental 
measurements of quantum systems having inhomogeneous potentials.  
These are common in the experimental arrangements used for precision 
measurements of quantum systems [3]; the Paul trap used in [1], for
example, constrains the ions within a harmonic oscillator potential.

Our model for the limited resolution of experimental measurements is
the simplest possible [13]:  the continuous spectrum of the operator
of interest is partitioned into congruent `bins' [16] corresponding 
to possible finite precision readings of a measuring device.  For the 
one dimensional position measurements with which we are concerned 
here, we take the bins to be intervals of length $\epsilon > 0$ and, 
in keeping with the computer science connotations of `finite 
precision', write $x = y\epsilon + z$ for a position $x \in \R$, with
$y \in \Z$ labelling the bin and $0 \le z < \epsilon$ the position 
inside the bin.  Resolving the mathematical difficulties in the usual
way [17], we take the set of position eigenstates $\{|x\rangle\}$ as
a basis for the Hilbert space $H$ of one particle wave functions.  Now
notice that writing 
$|x\rangle = |y\epsilon + z\rangle = |y\rangle \otimes |z\rangle$
defines a tensor product decomposition 
$H = H_{\rm coarse} \otimes H_{\rm fine}$.  Since we are taking the 
$H_{\rm fine}$ factor of the Hilbert space to be inaccessible to 
measurement, the situation is mathematically equivalent to the 
decomposition of Hilbert space into system and environment factors 
usual in discussions of decoherence (whether it is conceptualized as
damping [18], measurement [12,19,20], relaxation [21] or
dissipation [22]):  small scale position functions as an unmeasured 
`environment' `coupled' to the finite precision position `system'.

An arbitrary vector $|\psi\rangle \in H$ need not, of course, factor
as a tensor product of vectors in $H_{\rm coarse}$ and $H_{\rm fine}$.
Any quantum state represented as a density operator 
$\rho \in H \otimes H^*$, however, can be traced over $H_{\rm fine}$ 
to obtain a {\sl scale $\epsilon$ reduced density operator\/}
$\tilde\rho := {\rm Tr}_{\rm fine} \rho 
           \in H_{\rm coarse}^{\vphantom*} \otimes H_{\rm coarse}^*$,
describing the quantum state as measureable with finite precision.  
The trace over $H_{\rm fine}$ is exactly an integral over small 
scales:  for a pure state $|\psi\rangle \in H$, the density function
is $\rho(x,x') := \psi(x)\overline{\psi(x')}$ and the scale $\epsilon$ 
reduced density function is
$$
\tilde\rho(y,y') 
 := \int_0^{\epsilon} \psi(y\epsilon + z) 
                      \overline{\psi(y'\epsilon + z)}\,dz     \eqno(1)
$$
for $y,y' \in \Z$.  Notice that plane waves are tensor products 
relative to the decomposition of $H$ just as are the position 
eigenstates.  In this case the scale $\epsilon$ reduced density 
operator has rank 1:
$$
\tilde\rho(y,y') 
 = \int_0^{\epsilon} e^{ik(y\epsilon + z)} e^{-ik(y'\epsilon + z)}\,dz
 =: e^{ik(y-y')\epsilon} I(0),
$$
where $I(0) = \epsilon$; and thus the {\sl scale $\epsilon$ entropy\/} 
$S := -{\rm Tr}\tilde\rho\log\tilde\rho$ vanishes as it does for the
original pure state.

Now consider a superposition of plane waves:  
$\psi(x) = a_1 e^{ik_1x} + a_2 e^{ik_2x}$.  The scale $\epsilon$ 
reduced density function (1) is:
$$
\eqalign{
\tilde\rho(y,y') 
 &= a_1\bar a_1 e^{ik_1(y-y')\epsilon} I(0) +
    a_1\bar a_2 e^{i(k_1y-k_2y')\epsilon} I(k_1-k_2)               \cr
 &\quad +
    a_2\bar a_1 e^{i(k_2y-k_1y')\epsilon} I(k_2-k_1) +
    a_2\bar a_2 e^{ik_2(y-y')\epsilon} I(0),                       \cr
}
$$
where 
$$
I(k_1-k_2) 
 := \int_0^{\epsilon} e^{i(k_1-k_2)z}\,dz.                    \eqno(2)
$$
It is clear that in this case, for $k_1 \not= k_2$, $\tilde\rho$ has
rank 2.  In fact, $\tilde\rho$ is easy to diagonalize:  Its 
eigenvectors must be linear combinations 
$\phi(y') = b_1 e^{ik_1y'\epsilon} + b_2 e^{ik_2y'\epsilon}$; setting
$$
\sum_{y'} \tilde\rho(y,y') \phi(y') = \lambda \phi(y)
$$
we find that for $\phi$ to be an eigenvector $\lambda$ must be a root
of the characteristic equation
$$
0 = \det\pmatrix{      a_1\bar a_1 - \lambda & 
                 a_1\bar a_2 I(k_1-k_2)/I(0) \cr
                 a_2\bar a_1 I(k_2-k_1)/I(0) &
                       a_2\bar a_2 - \lambda \cr
                }.                                            \eqno(3)
$$
Now the entropy of the scale $\epsilon$ reduced density matrix does 
not vanish:  we can solve (3) for $\lambda$ and examine the dependence
of the eigenvalues (and thence the entropy) on $\epsilon$.  From (2) 
it is clear that the relevant quantity is $|k_1-k_2|\epsilon$:  when
$|k_1-k_2|\epsilon \ll 1$ the scale $\epsilon$ entropy almost 
vanishes; only when $\epsilon$ is at least comparable to 
$|k_1-k_2|^{-1}$, which we can consider to be the characteristic 
spatial scale of $\psi(x)$, is there substantial entropy.  As 
$\epsilon \to \infty$ the scale $\epsilon$ entropy (in bits) saturates
at 1.

Notice, furthermore, that the eigenvalues $\lambda$ depend on the 
Fourier coefficients $a_n$ in the superposition only through their 
squared norms $a_n\bar a_n$.  In fact, this is true for superpositions 
of any number of plane waves.  For a homogeneous potential plane waves 
are stationary states, so since the time evolution of each is by a 
phase $e^{-i\omega_nt}$, the scale $\epsilon$ entropy, even if it is
nonzero, must be constant in time; no initial state scale decoheres.

For an inhomogeneous potential the situation is more interesting:  the
eigenstates are still stationary, but they are not plane waves.  
Consequently the scale $\epsilon$ reduced density operator for a wave 
function expressed as a superposition of eigenstates is more 
complicated to diagonalize.  Generically the eigenvalues depend on the 
phases of the superposition coefficients as well as their squared 
norms and hence the scale $\epsilon$ entropy is time dependent.

Let us examine the case of the harmonic oscillator potential which is
present in the experimental arrangements described in [1,23,24] 
and proposed in [5], for example.  An initial eigenstate does not 
decohere, of course, since the density operator in this case is time 
independent.  So consider an initial superposition of two eigenstates 
$u_i$ of the harmonic oscillator Schr\"odinger operator, as was 
prepared in the experiments reported in [23,24]:
$\psi(x) = \bigl(u_1(x) - u_3(x)\bigr)/\sqrt{2}$, say.  The scale 
$\epsilon$ reduced density function $\tilde\rho(y,y')$ for such an 
superposition is a complicated and not particularly enlightening 
expression involving the error function, Erf$(\cdot)$.  More 
informative is direct numerical evaluation of the eigenvalues and the 
scale $\epsilon$ entropy as a function of time.

To do so we normalize the potential to $V(x) = x^2/2$ and calculate 
the reduced density matrix for $\epsilon = 1/2$.  (In order to 
preserve the symmetry of the problem we shift the domain of 
integration in (1) to $[-\epsilon/2,\epsilon/2)$.)  For numerical
purposes the wave function vanishes already for $|x|$ greater than
about 5 as is indicated by the probability distributions plotted in 
Figure~1; although $\epsilon = 1/2$ is a significant fraction of the
size of the support of the wave function it is still small relative to
the characteristic spatial scales, approximately 4 and 2, 
respectively, of $u_1$ and $u_3$, so we do not expect the scale 
$\epsilon$ entropy to be saturated.

\topinsert
$$
\epsfxsize=\halfwidth\epsfbox{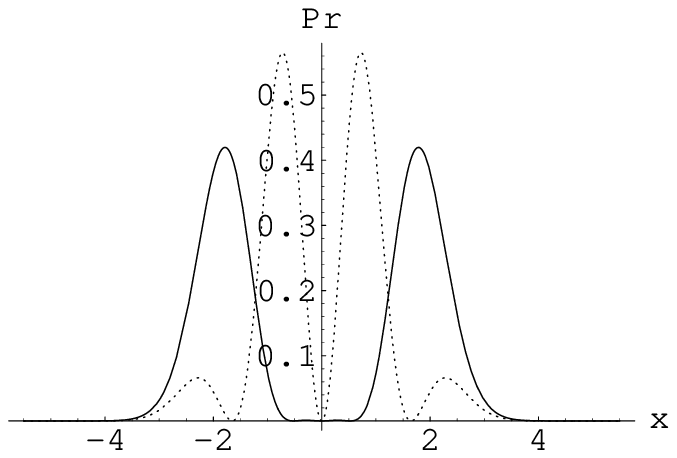}\hskip\chasm%
\epsfxsize=\halfwidth\epsfbox{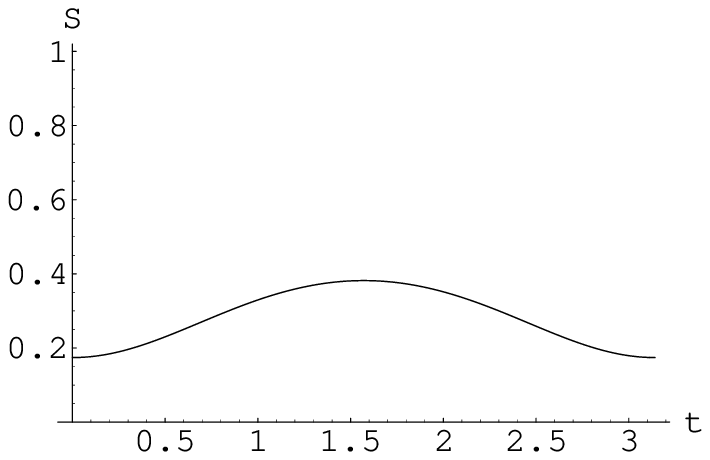}
$$
\hbox to\hsize{%
\vbox{\hsize=\halfwidth\eightpoint{%
\noindent{\bf Figure~1}.  The position probability distributions 
computed from the wave function at time 0 (solid curve) and time 
$\pi/2$ (dotted curve).
}}
\hfill%
\vbox{\hsize=\halfwidth\eightpoint{%
\noindent{\bf Figure~2}.  Entropy as a function of time for the scale
$\epsilon$ reduced density matrix of an initial superposition of 
harmonic oscillator eigenstates.
}}}
\endinsert

Figure~2 shows the scale $\epsilon$ entropy (in bits) computed for the 
time evolution of the initial superposition.  The scale $\epsilon$ 
entropy initially increases---the state as measured with finite 
precision decoheres.  In the longer run, however, the scale $\epsilon$ 
entropy is oscillatory.  In fact, since the eigenstates in the 
superposition have periods $4\pi/3$ and $4\pi/7$, respectively, the 
evolution must be exactly periodic with period $\pi$, as shown in 
Figure~2.  This type of oscillatory behavior for the entropy is 
familiar from studies of the decoherence of a finite dimensional 
quantum system coupled to a finite dimensional environment, \eg, one 
spin coupled to another [20,4].  Here the `environment' $H_{\rm fine}$ 
is infinite dimensional, but only a small finite dimensional 
subspace---determined by the Fourier components of the eigenstates
superposed in the initial state---has nonnegligible `coupling' to the 
essentially finite dimensional `system' $H_{\rm coarse}$.  This means, 
in addition, that the scale $\epsilon$ entropy is bounded above.  The 
maxima of the entropy curve occur at times when the wave function has 
the smallest characteristic spatial scale.  Figure~1 shows 
$|\psi(x,t)|^2$ for $t = 0$ and $t = \pi/2$, at which times the scale 
$\epsilon$ entropy is minimal and maximal, respectively.  

The minima of the entropy curve in Figure~2 do not give vanishing
entropy; that the entropy does not vanish at $t = 0$ (and would be 
even larger than it is were $\epsilon > 1/2$) is uncharacteristic of 
typical analyses of decoherence where the initial state is often taken
to factor relative to the tensor product decomposition of the Hilbert 
space [18--22,3,25].  To illustrate this situation for scale 
decoherence let us consider the case of a periodic potential where we 
can naturally take the initial state to be a plane wave which, as we 
noted earlier, is a tensor product relative to the scale decomposition 
of $H$.

Specifically, we consider the Schr\"odinger equation defined on the
circle $S^1 = \R/2\pi\Z$ with the potential
$$
V(x) = \cases{V_0 & $-\pi/2 \le x < \pi/2$ (mod $2\pi$);  \cr
              0   & otherwise.                            \cr
             }                                                \eqno(4)
$$
An eigenstate of energy $E$ is, of course, a linear combination of 
$e^{i\sqrt{E}x}$ and $e^{-i\sqrt{E}x}$ where $V(x) = 0$ and a linear
combination of $e^{i\sqrt{E-V_0}x}$ and $e^{-i\sqrt{E-V_0}x}$ where
$V(x) = V_0$.  The usual requirement that the wave function and its
first derivative be continuous determines the ratios of the 
coefficients in these superpositions.  Requiring the wave function to
be periodic makes the spectrum of eigenenergies discrete:  for 
$V_0 = 0$, $E_n = n^2$ for $n \in \Z_{\ge 0}$; for $V_0 > 0$ the
eigenvalues $E_n^{\pm}$ split (for $n > 0$) and are roots of a 
transcendental equation.  Let $\psi_n^{\pm}$ (and $\psi_0^+$) denote 
the corresponding symmetric/antisymmetric eigenstates.

When $V_0 > 0$ a plane wave $e^{ikx}$ for $k \in \Z$, although
periodic, is not an eigenstate.  Rather it is a superposition of
eigenstates with dominant contributions coming from ones with energies
near $k^2$ and near $k^2 + V_0$.  Just as in our earlier example of a 
superposition of two harmonic oscillator eigenstates, this means that
the initial plane wave state will decohere.  To make the scales
comparable, consider the case $k = 1$ in the potential (4) with 
$V_0 = 3$.  We expect the dominant components of the superposition to
be eigenstates with energies near 1 and near 4.  In fact,
$$
\eqalign{
e^{ix} \approx 
 &-0.41 \psi_0^+ +0.55 \psi_1^+ +0.66 i \psi_1^- +0.18 \psi_2^+ 
  +0.24 \psi_2^-                                                   \cr
 &+0.01 \psi_3^+ +0.03   \psi_3^- -0.02 \psi_4^-,                  \cr
}                                                             \eqno(5)
$$
where all components with amplitudes of norm at least 0.0001 are 
included and the energies of the first 5 eigenstates listed are 
approximately 0.52, 3.27, 2.01, 5.41 and 5.84, respectively, all of 
which are reasonably close to 1 and/or 4.  The corresponding 
characteristic spatial scales are near $2\pi$ and $\pi$, respectively,
so we set the precision $\epsilon = \pi/4$ to be comparable to the 
$\epsilon = 1/2$ precision we used in the harmonic oscillator 
calculation.

Figure~3 shows the scale $\epsilon$ entropy (in bits) for the $k = 1$
initial plane wave (5).  Since this initial state does factor as a 
tensor product relative to the scale decomposition of $H$, the scale 
$\epsilon$ entropy vanishes at $t = 0$.  Then, just as in Figure~2,
the scale $\epsilon$ entropy increases (to a similar maximum value)
before decreasing.  In this case, however, the multiple higher 
frequency components in the original superposition manifest themselves
in the irregularity of the entropy curve and prevent the entropy from
decreasing all the way to its initial (zero) value before it again
increases.  

\topinsert
$$
\epsfxsize=\halfwidth\epsfbox{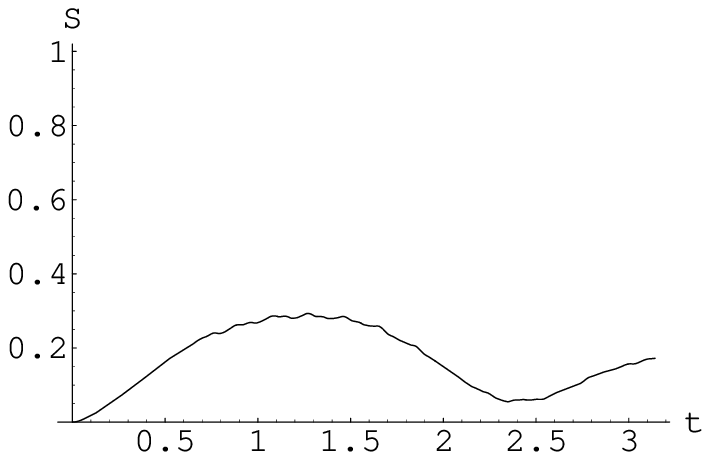}\hskip\chasm%
\epsfxsize=\halfwidth\epsfbox{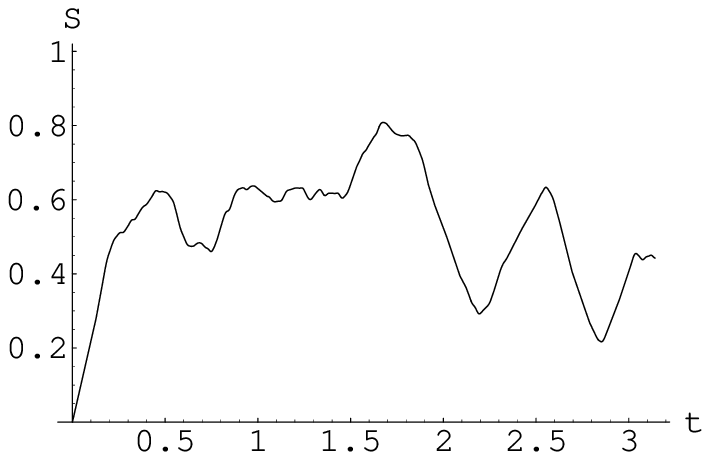}
$$
\hbox to\hsize{%
\vbox{\hsize=\halfwidth\eightpoint{%
\noindent{\bf Figure~3}.  Entropy as a function of time for the scale
$\epsilon$ reduced density matrix of an initial $k = 1$ plane wave in
the potential (4) with $V_0 = 3$.
}}
\hfill%
\vbox{\hsize=\halfwidth\eightpoint{%
\noindent{\bf Figure~4}.  Entropy as a function of time for the scale
$\epsilon$ reduced density matrix of an initial $k = 1$ plane wave in
the potential (4) with $V_0 = 15$.
}}}
\endinsert

Measuring position with greater precision, \ie, decreasing $\epsilon$,
reduces the amount of scale $\epsilon$ decoherence.  On the other 
hand, increasing the amplitudes of higher frequency components
amplifies the scale $\epsilon$ decoherence.  Figure~4 shows the scale
$\epsilon$ entropy for the potential (4) with $V_0 = 15$ but still 
with $\epsilon = \pi/4$ and initial state a plane wave with $k = 1$ as 
in Figure~3.  Now the dominant components are eigenstates with 
energies near 1 and 16, and with characteristic spatial scales near 
$2\pi$ and $\pi/2$, respectively.  Including only components of the 
same relevance as in (5) we have now
$$
\eqalign{
e^{ix} \approx
 &-0.49  \psi_0^+ +0.11 \psi_1^+ +0.54i \psi_1^- +0.44 \psi_2^+ 
  +0.29i \psi_2^-                                                   \cr
 &+0.25  \psi_3^+ +0.36 \psi_3^- -0.01  \psi_4^- -0.01 \psi_5^+ 
  -0.06  \psi_5^-                                                   \cr
 &+0.02  \psi_6^- +0.02 \psi_7^- -0.01  \psi_8^-,                   \cr
}                                                              \eqno(6)
$$
where the energies of the first 7 eigenstates listed are approximately
0.74, 6.47, 2.92, 15.28, 11.15, 17.49 and 18.47, respectively, all of 
which are near the range $[1,16]$.  Now the scale $\epsilon$ entropy 
increases even more rapidly from 0 than in Figure~3, and the presence 
of additional high frequency components makes the entropy curve so 
irregular as to obscure {\sl any\/} approximate periodicity in time.  
The common period of the eigenstates with appreciable amplitudes in (6) 
is so large that the initial state undergoes essentially irreversible 
scale $\epsilon$ decoherence.

These last two computations demonstrate that initial tensor product
states scale decohere with the same `ultraviolet burst' that is 
familiar from other decoherence calculations [3].  Moreover, the last
computation shows that when the finite precision `system' `couples'
strongly to multiple dimensions of the small scale `environment', 
scale decoherence also approaches the irreversible behavior of 
decoherence by an infinite dimensional thermal environment [18--22].

Most importantly, we have shown that scale decoherence occurs whenever 
the potential is inhomogeneous, which includes the experimentally 
relevant situation of a harmonic oscillator,%
\sfootnote*{Eigenstates, as we have seen, do not scale decohere.  Nor, 
            in a harmonic oscillator potential, does a coherent 
            gaussian wave packet since evolution preserves its shape;
            a superposition of such packets will, however, scale 
            decohere.}
and is mathematically identical to any other form of decoherence.
Thus scale decoherence is a real experimental effect for finite 
precision measurements.  Although the measurements described in [1] 
are not direct position measurements, the demonstration of mesoscopic
superpositions depends on the measurement of spin down probability as
a function of phase separation $\phi \in [-\pi,\pi)$.  At least part
of the unexplained decoherence [24] in the measurements for 
initially superposed harmonic oscillator coherent states 
$|\alpha\rangle$ and $|-\alpha\rangle$ with $\alpha \approx 2.97(6)$ 
may be scale decoherence due to finite precision measurement of the
relative phase angle $\phi$---this is consistent with the recognized
fluctuations in $\phi$ [1] and the finer scale variation of the 
signal (as a function of $\phi$) in this case than for smaller values 
of $\alpha$.  Similarly, we may expect scale decoherence in the larger
$\alpha$ cavity QED experiments proposed in [2].

Finally, we remark that scale decoherence affects the measurement and
phase space tomographic reconstruction of the density matrix (or
equivalently, the Wigner function) for a quantum state [5,23,24].
While not originally conceptualized as decoherence, it is an old
result [26], revisited in the consistent histories program [27], 
that smearing the Wigner function over a suitable volume of phase 
space produces a positive probability distribution.  Finite precision
measurement effects exactly such a smearing, although for sufficiently
small $\epsilon$ the resulting scale $\epsilon$ decoherence does not
eliminate all negative values of the Wigner function---and the limited
precision need not originate in the quantum mechanical uncertainty of
simultaneous position and momentum measurements.  Since reconstruction 
of the density matrix/Wigner function is now often performed in 
quantum optics [28] as well as in atomic physics, it may prove useful 
to consider optical homodyne/phase space tomography based on 
incomplete data [29] from the perspective of scale decoherence.

\noindent{\bf Acknowledgements}
\nobreak

\nobreak
\noindent I thank Hideo Mabuchi for his thoughtful comments on a 
preliminary version of this paper and Sun Microsystems for providing 
computational support.

\global\setbox1=\hbox{[00]\enspace}
\parindent=\wd1

\noindent{\bf References}
\bigskip

\parskip=0pt
\item{[1]}
C. Monroe, D. M. Meekhof, B. E. King and D. J. Wineland,
``A `Schr\"odinger cat' superposition state of an atom'',
\Sc\ {\bf 272} (1996) 1131--1136.

\item{[2]}
M. Brune, E. Hagley, J. Dreyer, X. Ma\^{\i}tre, A. Maali, 
C. Wunderlich, J. M. Raimond and S. Haroche,
``Observing the progressive decoherence of the `meter' in a quantum
  measurement'',
\PRL\ {\bf 77} (1996) 4887--4890.

\item{[3]}
J. R. Anglin, J. P. Paz and W. H. Zurek,
``Deconstructing decoherence'',
\PRA\ {\bf 55} (1997) 4041--4053.

\item{[4]}
C. H. Woo,
``Linear and nonlinear evolution of reduced density matrices'',
\PLA\ {\bf 217} (1996) 43--46.

\item{[5]}
M. Tegmark,
``Measuring quantum states:  Experimental setup for measuring the
  spatial density matrix'',
\PRA\ {\bf 54} (1996) 2703--2706;\hfb
L. G. Lutterbach and L. Davidovich,
``Method for direct measurement of the Wigner function in cavity
  QED and ion traps'',
\PRL\ {\bf 78} (1997) 2547--2550.

\item{[6]}
D. P. DiVincenzo,
``Quantum computation'',
\Sc\ {\bf 270} (1995) 255--261;\hfb
A. Barenco and A. Ekert,
``Quantum computation'',
\APS\ {\bf 45} (1995) 1--12;\hfb
and references therein.

\item{[7]}
\deutsch\ and R. Jozsa,
``Rapid solution of problems by quantum computation'',
\PRSLA\ {\bf 439} (1992) 553--558;\hfb
D. R. Simon,
``On the power of quantum computation'',
in S. Goldwasser, ed.,
{\sl Proceedings of the 35th Symposium on Foundations of Computer 
Science}, Santa Fe, NM, 20--22 November 1994
(Los Alamitos, CA:  IEEE Computer Society Press 1994) 116--123.

\item{[8]}
E. Bernstein and U. Vazirani,
``Quantum complexity theory'',
in {\sl Proceedings of the 25th ACM Symposium on Theory of Computing},
San Diego, CA, 16--18 May 1993
(New York:  ACM Press 1993) 11--20;\hfb
P. W. Shor,
``Algorithms for quantum computation:  discrete logarithms and 
  factoring'',
in S. Goldwasser, ed.,
{\sl Proceedings of the 35th Symposium on Foundations of Computer 
Science}, Santa Fe, NM, 20--22 November 1994
(Los Alamitos, CA:  IEEE Computer Society Press 1994) 124--134.

\item{[9]}
J. I. Cirac and P. Zoller,
``Quantum computation with cold trapped ions'',
\PRL\ {\bf 74} (1995) 4091--4094;\hfb
Q. A. Turchette, C. J. Hood. W. Lange, H. Mabuchi and H. J. Kimble,
``Measurement of conditional phase shifts for quantum logic'',
\PRL\ {\bf 75} (1995) 4710--4713.

\item{[10]}
W. G. Unruh, 
``Maintaining coherence in quantum computers'',
\PRA\ {\bf 51} (1995) 992--997;\hfb
I. L. Chuang, R. Laflamme, P. Shor and W. H. Zurek,
``Quantum computers, factoring and decoherence'',
\Sc\ {\bf 270} (1995) 1633--1635;\hfb
G. M. Palma, K.-A. Souminen and A. Ekert,
``Quantum computers and dissipation'',
\PRSLA\ {\bf 452} (1996) 567--584;\hfb
C. Miquel, J. P. Paz and R. Perazzo,
``Factoring in a dissipative quantum computer'',
\PRA\ {\bf 54} (1996) 2605--2613.

\item{[11]}
P. W. Shor,
``Scheme for reducing decoherence in quantum memory'',
\PRA\ {\bf 52} (1995) R2493--R2496;\hfb
A. R. Calderbank and P. W. Shor,
``Good quantum error-correcting codes exist'',
\PRA\ {\bf 54} (1996) 1098--1105;\hfb
and references therein.

\item{[12]}
P. W. Shor,
``Fault-tolerant quantum computation'',
in
{\sl Proceedings of the 37th Symposium on Foundations of
  Computing}, Burlington, VT, 14--16 October 1996
(Los Alamitos, CA:  IEEE Computer Society Press 1996) 56--65;\hfb
D. Gottesman,
``Theory of fault-tolerant quantum computation'',
\PRA\ {\bf 57} (1998) 127--137.

\item{[13]}
J. von Neumann,
{\it Mathematische Grundlagen der Quantenmechanik\/}
(Berlin:  Spring\-er-Verlag 1932); 
transl.\ by R. T. Beyer as
{\sl Mathematical Foundations of Quantum Mechanics\/}
(Princeton:  Princeton University Press 1955).

\item{[14]}
C. M. Caves,
``Quantum mechanics of measurements distributed in time.  
  A path-integral formulation'',
\PRD\ {\bf 33} (1986) 1643--1665.

\item{[15]}
J. B. Hartle,
``Spacetime coarse grainings in nonrelativistic quantum
  mechanics'',
\PRD\ {\bf 44} (1991) 3173--3196;\hfb
H. F. Dowker and J. J. Halliwell,
``Quantum mechanics of history:  The decoherence functional in
  quantum mechanics'',
\PRD\ {\bf 46} (1992) 1580--1609.

\item{[16]}
M. H. Partovi,
``Entropic formulation of uncertainty for quantum measurements'',
\PRL\ {\bf 50} (1983) 1883--1885.

\item{[17]}
See, \eg,
A. Messiah, 
{\sl Quantum Mechanics\/} Vol.\ I,
transl.\ from the French by G. M. Temmer
(New York:  John Wiley \& Sons 1958) Appendix A.

\item{[18]}
L. Landau,
``{\it Das D\"ampfungsproblem in der Wellenmechanik\/}'',
\ZP\ {\bf 45} (1927) 430--441;\hfb
L. Di\'osi,
``Landau's density matrix in quantum electrodynamics'',
\FP\ {\bf 20} (1990) 63--70.

\item{[19]}
F. London and E. Bauer,
``{\it La th\'eorie de l'observation en m\'echanique quantique\/}'',
No.\ 775 of {\sl Actualit\'es scientifiques et industrielles:  
  Expos\'es de physique g\'en\'erale}, 
{\it publi\'es sous la direction de Paul Langevin\/}
(Paris:  Hermann 1939); 
transl.\ by A. Shimony, by J. A. Wheeler and W. H. Zurek, 
  and by J. McGrath and S. McLean McGrath, reconciled in
J. A. Wheeler and W. H. Zurek,
{\sl Quantum Theory and Measurement}
(Princeton:  Princeton University Press 1983) 217--259.

\item{[20]}
\zurek,
``Pointer basis of quantum apparatus:  Into what mixture does the wave
  packet collapse?'',
\PRD\ {\bf 24} (1981) 1516--1525;\hfb
\zurek,
``Environment-induced superselection rules'',
\PRD\ {\bf 26} (1982) 1862--1880;\hfb
E. Joos and H. D. Zeh,
``The emergence of classical properties through interaction with the
  environment'',
\ZPB\ {\bf 59} (1985) 223--243.

\vfill\eject

\item{[21]}
R. K. Wangness and F. Bloch,
``The dynamical theory of nuclear induction'',
\PR\ {\bf 89} (1953) 728--739;\hfb
F. Block,
``Generalized theory of relaxation'',
\PR\ {\bf 105} (1957) 1206--1222;\hfb
W. H. Louisell and L. R. Walker,
``Density-operator theory of harmonic oscillator relaxation'',
\PR\ {\bf 137} (1965) B204--B211.

\item{[22]}
I. R. Senitzky,
``Dissipation in quantum mechanics.  The harmonic oscillator'',
\PR\ {\bf 119} (1960) 670--679.

\item{[23]}
D. Leibfried, D. M. Meekhof, B. E. King, C. Monroe, W. M. Itano and
D. J. Wineland,
``Experimental determination of the motional quantum state of a 
  trapped atom'',
\PRL\ {\bf 77} (1996) 4281--4285.

\item{[24]}
W. M. Itano, C. Monroe, D. M. Meekhof, D. Liebfried, B. E. King and
D. J. Wineland,
``Quantum harmonic oscillator state synthesis and analysis'',
in 
M. G. Prentiss and W. D. Phillips, eds.,
{\sl Atom Optics}, San Jose, CA, 10--12 February 1997,
{\sl Proc.\ SPIE\/} {\bf 2995} (1997) 43--55.

\item{[25]}
\dajm,
``Decoherence in the Dirac equation'',
quant-ph/9804023.

\item{[26]}
K. Husimi,
``Some formal properties of the density matrix'',
\PPMSJ\ {\bf 22} (1940) 264--314;\hfb
N. D. Cartwright,
``A non-negative Wigner-type distribution'',
\PA\ {\bf 83} (1976) 210--212;\hfb
R. E. O'Connell and E. P. Wigner,
``Some properties of a non-negative quantum-mechanical distribution
  function'',
\PLA\ {\bf 85} (1981) 121--126.

\item{[27]}
J. J. Halliwell,
``Smeared Wigner functions and quantum-mechanical histories'',
\PRD\ {\bf 46} (1992) 1610--1615.

\item{[28]}
K. Vogel and H. Risken,
``Determination of quasiprobability distributions in terms of 
  probability distributions for the rotated quadrature plane'',
\PRA\ {\bf 40} (1989) 2847--2849;\hfb
D. T. Smithey, M. Beck, M. G. Raymer and A. Faridani,
``Measurement of the Wigner distribution and the density matrix
  of a light mode using optical homodyne tomography:  Application
  to squeezed states and the vacuum'',
\PRL\ {\bf 70} (1993) 1244--1247;\hfb
W. Vogel and D.-G. Welsch,
{\sl Lectures on Quantum Optics\/}
(Berlin:  Akademie-Verlag 1994) Chapters 3 and 6;\hfb
and references therein.

\item{[29]}
V. Bu\v{z}ek, G. Adam and G. Drobn\'y,
``Reconstruction of Wigner functions on different observation 
  levels'',
\AP\ {\bf 245} (1996) 37--97;\hfb
P. T\"orm\"a,
``Finite number of measurements in optical homodyne tomography'',
\JMO\ {\bf 43} (1996) 2437--2447;\hfb
R. Derka, V. Bu\v{z}ek, G. Adam and P. L. Knight,
``From quantum Bayesian inference to quantum tomography'',
quant-ph/9701029;\hfb
V. Bu\v{z}ek, G. Drobn\'y, G. Adam, R. Derka, and P. L. Knight,
``Reconstruction of quantum states of spin systems via the Jaynes
  principle of maximum entropy'',
\JMO\ {\bf 44} (1997) 2607--2627.

\bye